\documentclass{rpvol10x7}

\begin{document}







\toctitle{Tullio Regge's legacy: Regge calculus and discrete (quantum) gravity}
\tocauthor{Ruth M. Williams}
\tocsource{{\it Source} {\bf vol}, page (year)}

\chapter{Tullio Regge's legacy: Regge calculus and discrete (quantum) gravity}

\author{John W. Barrett} 
\address{School of Mathematical Sciences, University Park, Nottingham
  NG7 2RD, UK, EU}

\author{Daniele Oriti}
\address{Max Planck Institute for Gravitational Physics (Albert
  Einstein Institute), Am M\"uhlenberg 1, D-14476 Golm, Germany, EU, \\ 
II Institute for Theoretical Physics, University of
  Hamburg, Luruper Chaussee 149, D-22761 Hamburg, Germany, EU and \\ 
 Arnold-Sommerfeld-Center for Theoretical
Physics, Ludwig-Maximilians-Universit\"at, Theresienstrasse 37, D-80333
M\"unchen, Germany, EU}

\author{Ruth M. Williams}
\address{Girton College, Cambridge CB3 0JG, UK, EU and \\ 
DAMTP, CMS, Wilberforce Road, Cambridge CB3 0WA, UK. EU}

\vspace{2cm}

\noindent
The review paper \lq\lq Discrete Structures in Physics"\cite{RW}, written in
2000, describes how Regge's discretization of Einstein's theory has been
applied in classical relativity and quantum gravity. Here,
developments since 2000 are reviewed briefly, with
particular emphasis on progress in quantum gravity through spin foam models
and group field theories.

\section{Introduction}

Regge's seminal 1961 paper, \lq\lq General Relativity without
Coordinates"\cite{TR}, is a brief and rigorous description of how to
approximate curved spaces and space-times by simplicial complexes, with the
curvature distributed on simplicial nets of co-dimension 2. This was
motivated by the desire to model complicated topologies and to obtain a
deeper geometrical insight. The action for a simplicial space was written
down and the analogue of the Einstein equations derived, making use of the
Schl\"afli identity. In the last sentence, Regge mentions that the approach
could be useful in numerical work. The paper shows clearly Regge's profound
geometrical intuition; he admitted later that, having seen how to
triangulate space-time manifolds, he was really interested in triangulating
group manifolds. His approach became known as Regge calculus.

The paper opened up whole new areas of research. At first it was used for
calculations of the evolution of model universes in classical general 
relativity, and later
in efforts to formulate a theory of simplicial quantum gravity. Regge
himself never worked extensively in these fields, but maintained an
interest and in fact made a connection which led to one of the most
important approaches to discrete quantum gravity.

In the classical theory, in an unpublished paper with Lund,
\lq\lq Simplicial Approximation to some Homogeneous Cosmologies"\cite{LR},
Regge contributed to the continuous time three-plus-one formulation of
Regge calculus by writing down the form of the action for homogeneous and
isotropic spaces, giving an explicit form for the Hamiltonian constraint
for such spaces. (The momentum constraints are identically satisfied in
this case, but finding a simplicial form for them is more challenging.)

The paper which proved so influential in simplicial quantum gravity was
written with Ponzano in 1968, and entitled \lq\lq Semiclassical limit of
Racah coefficients"\cite{PR}, these coefficients being a very useful
tool in the calculation of
matrix elements in atomic physics. Almost in passing, the authors point out
the relationship between a sum involving the asymptotic values of
6j-symbols associated with a triangulated three-manifold, and the path
integral for three-dimensional simplicial gravity with the Regge action.
This work was largely neglected until the 1990s when
mathematicians\cite{TV1} were writing down invariants of three-manifolds,
hoping that they would help in the classification, and it was realised that
there was a very close connection with the work of Ponzano and Regge. In
fact the the Turaev-Viro model appeared to be a regularised version of the
Ponzano-Regge model.

There were many attempts to generalise this work to four dimensions,
in the general language of spin foam models, the first being the 
Barrett-Crane model, which will be discussed in a later section. This,
in its turn, renewed interest in group field theories, first introduced
for the Ponzano-Regge-Turaev-Viro models, which are described in the 
penultimate section.

Regge was clearly still thinking about discretization, and sometime in the
1970s produced a draft paper entitled \lq\lq Discrete Yang-Mills theories".
(He gave his address as  \lq The New Jersey Mental Hospital for Retired
Physicists and Compulsory Psychopaths' - in mirror writing on one copy; he
was at the Institute for Advanced Study in Princeton at the time!) A final
version was published in the Festschrift for Yuval Ne'eman\cite{TRN}.

For the millenium edition of the Journal of Mathematical Physics, Regge
co-authored a review paper\cite{RW} on Discrete Structures in Physics. This
describes in detail forty years of development of Regge calculus and
formulations of quantum gravity which have grown out of it. (When the
review was being written, Regge was fascinated by computer-generated art
work, and produced a picture he called \lq\lq Johann Sebastian Beach", with
piano keys forming the sea wall!) This article will not attempt to cover
the material in that review, but rather to describe some of the
developments since it was written, under the headings Classical Formalism,
Numerical Relativity, Simplicial Quantum Gravity, the Barrett-Crane Model
and Group Field Theories.

\section{Classical Formalism}

Many of the developments in classical Regge calculus since 2000 have come
from Warner Miller and his collaborators. Gentle and Miller\cite{GM1}
developed an algorithm which produced spacelike simplicial hypersurfaces
with constant mean curvature. The equations resulting from a variational
approach to this problem were solved together with the Regge equations, and
a formulation was proposed which was compatible with Sorkin 
evolutions\cite{BGM}.
McDonald and Miller\cite{MM} have emphasised the important role that the
dual lattice plays in Regge calculus. They derived a vertex-based scalar
curvature using a new lattice obtained both from the simplicial lattice and
its dual; this was a vertex-based weighted average of deficit angles per
weighted average of dual areas. Mean curvature on a simplicial lattice was
also discussed by Conbove, Miller and Ray\cite{CMR}.

Turning their attention to non-vacuum space-times, Gentle, Kheyfets,
McDonald and Miller\cite{GKMM} derived a conservation law in Regge calculus
by equating the discrete Bianchi identity to a sum of components of the
stress-energy tensor projected along the edges of the simplicial lattice.
In principle, this extends Regge calculus in a natural way to non-vacuum
space-times, provided an appropriate form of the stress-energy tensor can
be written down. McDonald and Miller\cite{MM2} considered the coupling of
non-gravitational fields to simplicial space-times, and constructed the
lattice action for scalar fields, the Maxwell field tensor and Dirac
particles, using discrete differential forms.

An exact form for the Bianchi identity on a simplicial lattice has been
given by Hamber and Kegel\cite{HK}. It is valid for arbitrarily curved
manifolds, but is not linear in the curvatures in general.

Arjwahjoedi and Zen\cite{AZ} have focussed on (2+1)-Regge calculus and
obtained expressions for discrete curvatures, the Bianchi identity and the
Gauss-Codazzi equation, in an attempt to give geometrical clarification to
earlier work. It was shown that the standard formulae for these can be
obtained in the continuum limit. The main result is that the Gauss-Codazzi
equation is very closely related to the dihedral angle formula, which
relates an $n$-dimensional angle between two $(n-1)$-dimensional simplices, to
$(n-1)$-dimensional angles between two $(n-2)$-dimensional simplices. This work
should now be extended to 3+1 dimensions.

In a contrasting approach, H\"ohn\cite{PH} has studied canonical Regge
calculus expanded to linear order about a flat background. He showed how to
use the Pachner moves in the evolution of a spacelike hypersurface and
identified the gauge and \lq graviton' degrees of freedom. The constraints
generating the vertex displacement symmetry are consistent with the
dynamics and are preserved by the Pachner moves. However it seems that the
gauge symmetries will be broken in higher order approximations.

\lq Area Regge calculus\rq ~is an alternative approach to Regge's scheme, first
suggested by Rovelli\cite{CR} and motivated by loop quantum gravity and
spinfoam models. In this, the triangle areas rather than the edge lengths
are the fundamental variables. The variational principle leads to vanishing
deficit angles, and metric discontinuities. Wainwright and
Williams\cite{WW} showed how a simple class of geometries with a
discontinuity across a hyperplane leads to refractive wave geometries,
which are generalised solutions of general relativity\cite{JB} .

Neiman\cite{YN} posed the question of whether \lq area Regge calculus\rq ~is a
valid discretisation of general relativity. It was found that in cases of
interest (Euclidean, or Lorentzian with spacelike tetrahedra) the
distributional scalar curvature is non-zero and has the same sign round all
triangles, so cannot average to zero, which would be required in general
relativity. A non-zero cosmological constant does not solve the problem. If
all tetrahedra were null, the argument would not hold, but it is
combinatorically impossible to triangulate space-time with null tetrahedra.
A carefully-constructed space-time with both timelike and spacelike
tetrahedra might solve the problem but that seems unlikely. Neiman's result
has implications for the Barrett-Crane model.

Dittrich and Speziale\cite{DS} introduced a modified version of Regge 
calculus in four dimensions where the fundamental variables are areas and a 
certain class of angles. The constraints are local on the triangulation, so 
this formulation solves the long-standing problem of how to implement the 
area constraints.

In a series of papers, Bahr and Dittrich\cite{BD} constructed a further
interesting variation on conventional Regge calculus. Motivated by the fact
that diffeomorphism symmetry is broken if the simplicial space-time is not
flat, which leads to pseudo-constraints in the canonical theory, they
considered the replacement of flat simplices by ones with constant
sectional curvature (consistent with the presence of a cosmological
constant). The lengths were replaced by dihedral angles as the basic
variables and there were no constraints in the first order formalism.

Christiansen\cite{SC} studied the convergence of linearised
three-dimensional Regge calculus, and showed that the first non-trivial
terms in the Regge action (the second variations) agree with the
Einstein-Hilbert action.

Regge calculus as originally formulated is a torsion-free theory. Schmidt
and Kohler\cite{SK} have generalised it to include dislocations in the
simplicial lattice, corresponding to torsion singularities. Similarly to
curvature, torsion is distributed on the simplices of co-dimension two, and
in four dimensions the contribution to the action (a discrete version of
the Einsten-Cartan action) involves the square of $b_{(i)}$, the part of
the Burgers vector parallel to the triangle. The variables of the theory
are taken to be not only the edge lengths but also the $b_{(i)}$, and the
field equations show that the Burgers vector couples algebraically to the
matter term. Thus the torsion vanishes in the absence of matter. See also
work by Drummond\cite{ITD} and Barrett \cite{Barrett-torsion} on torsion in Regge calculus and work by
Xue\cite{SX} on the Einstein-Cartan theory in the quantum Regge calculus
context.

Teleparallel gravity is a version of gravity where the curvature vanishes
but the torsion is non-zero and acts as a force. Pereira and
Vargas\cite{PV} formulated the teleparallel version of Regge calculus, with
the action again proportional to the square of the Burgers vector parallel
to the triangular hinge. The variables are taken to be just the edge
lengths and, in the variation, the Burgers vector is treated as constant. The
result is that in this theory, the Burgers vector does not necessarily
vanish in the absence of matter and torsion is a propagating field.

\section{Numerical Relativity}

Many of the applications of Regge calculus in numerical relativity have
been relatively small-scale calculations, where the space-time has
considerable symmetry and where, in many cases, the continuum solution is
known. Examples of such work will be described first, then there will be
brief more general comments about Regge calculus as a tool in numerical
relativity.

A suitable cosmology for testing Regge calculus is the Kasner model, which
was first examined in this way by Lewis\cite{SL}, who was unable to obtain
the full set of Kasner-Einstein equations in the continuum limit.
Gentle\cite{AG1} re-examined the problem and discovered that Lewis had
neglected one type of curvature. With this included, the full set of
equations was derived in the continuum limit and accurate numerical
solutions obtained. Brewin and Gentle\cite{BG}, again in the context of the
Kasner cosmology, investigated how the simplicial solutions were
second-order accurate approximations to the continuum solution, but the
residual of the Regge equations evaluated on the continuum solution did not
converge, because of a wave-like disturbance with high frequency but low
amplitude in the simplicial solution. The amplitude of this wave converged
to zero as the discretisation was refined, and it did not affect the
overall second-order accuracy of the simplicial solution. Brewin\cite{LB}
also looked at evolution of the Kasner cosmology using both Regge calculus
and his smooth lattice method; both produced convergent approximations to
the exact solution, but Regge calculus was two orders of magnitude slower.

Closed FLRW universes with positive cosmological constant were considered
in three dimensions by Tsuda and Fujiwara\cite{TF}. The Cauchy surfaces
were taken to be the surfaces of the regular polyhedral solids in three
dimensions, with prisms connecting the surfaces at subsequent times. The
numerical solution was found to deviate from the continuum solution at
large times. The triangles on the Cauchy surface were then subdivided and
the new vertices projected onto a sphere to give a geodesic dome. It was
found that the numerical solution agreed with the continuum one in the
limit of infinite subdivision. This work should now be extended to four
dimensions, and matter included.

In a series of papers, Liu and Williams discussed various aspects of the
time evolution of simple model universes. In the context of a closed empty
universe with positive cosmological constant, they investigated whether to
apply the variational principle to the action then impose symmetries (local
variation) or to impose symmetries then vary (global variation). It emerged
that local variation does not generally lead to a viable set of Regge
equations\cite{LW1}. They then considered lattice universes with point
masses distributed on a regular lattice on the Cauchy surface. Constraints
were obtained on the distribution of the masses for the model to be stable.
The evolution resembled that of a closed FLRW dust-filled universe. When
one mass was perturbed, the model's evolution was well-behaved, with the
expansion increasing in magnitude as the perturbation increased\cite{LW2}.

Progress with the classical evolution of a spacelike hypersurface was made
in the early 1990s when it was realised that, in general, the Regge
equations decouple into a collection of much smaller groups, leading to the
so-called \lq Sorkin evolution'\cite{BGM}, already mentioned. De Felice
and Fabri\cite{DFF}
showed that the vertices in the 600-tetrahedron tessellation of the
three-sphere fall into {\it five} distinct classes, not four as in the
original paper on the Parallelisable Implicit Evolution
Scheme\cite{BGM}. They also investigated the so-called \lq stopping point'
which has plagued many numerical calculations of the evolution of model
universes using Regge calculus; this point is where the evolution stops
well before the spatial volume becomes zero. Brewin\cite{LB1} had suggested
that this occurs when the \lq vertical' edges become spacelike as the
evolution is so fast. De Felice and Fabri showed that it is a
causality-breaking singularity in the effective metric in Regge calculus in
such calculations. The issue of causality in this approach was also
investigated by Khavari\cite{PK}, who generalised the triangle inequalities
in Euclidean space to inequalities on edges of triangles in Minkowski
space. When causality is included in this way in the Parallelisable
Implicit Evolution Scheme, the \lq stopping point' problem is resolved,
which is significant progress. The revised algorithm was applied
successfully to the FLRW universe\cite{KD}. Khavari also set up a Regge
calculus version of the Raychaudhuri equation\cite{PK}. In both 2+1 and 3+1
dimensions, analogues for the average expansion and the shear scalar were
found.

More generally, in a brief review, Gentle and Miller\cite{GM2} set out a
programme for large-scale numerical work using Regge calculus and progress
was discussed by Gentle\cite{AG2}. In the spherically symmetric case, the
static Schwarzschild solution was modelled to high accuracy. Axisymmetric
initial data was constructed for Brill waves, and for a black hole with
Brill wave perturbations. Regge calculus methods produced a very accurate
approximation to Misner's analytic solution for initial data for the
head-on collision of two equal mass non-rotating black holes. The generic
axisymmetric code developed by Gentle, Miller and collaborators was used to
obtain the time evolution of the Brill wave initial data, the first
successful time evolution of gravitational radiation on a lattice. These
successes are very encouraging but there is still much work to be done,
particularly on the inclusion of matter and the relation of lattice
approaches to standard finite element methods for solving differential
equations. Because of the investment of time and computer power needed to
develop Regge lattice methods further, it seems that current work in
numerical relativity relies more on the very successful standard finite
difference techniques to perform calculations in astrophysics, including
the recent focus on gravitational waves.

\section{Simplicial quantum gravity}

This section reports on some of the recent work in quantum Regge calculus,
but excludes the Barrett-Crane model and group field theories, which are
discussed separately.

In a new approach to quantum Regge calculus, Xue\cite{SX} translated the
Einstein-Cartan theory to a Euclidean lattice, with the tetrad field, the
gauge field and the spin connection assigned to the edges. The partition
function and effective action are constructed, and the vacuum expectation
values of diffeomorphism and local gauge-invariant quantities can in
principle be calculated. Some calculations in two dimensions are presented.

In a paper dedicated to Rafael Sorkin, one of the pioneers of Regge
calculus, Gambini and Pullin\cite{GP} apply their \lq consistent
discretization' approach to Regge calculus. They set up a canonical
formalism which is free of constraints, with the time evolution equivalent
to a canonical transformation. Some of the edge lengths are effectively
Lagrange multipliers and can be eliminated using their equations of motion.
This seems to avoid the common problem of \lq spikes' (thin, elongated
simplices) in the Lorentzian case. Quantization is achieved by writing down
a path integral with the \lq Lagrange multipliers' eliminated; the
measure is determined uniquely by the unitary transformation that
implements the dynamics. It is straightforward to include topology change
in this framework.

Work by Dittrich and H\"ohn\cite{DH} generalizes the \lq consistent 
discretization' approach to allow for arbitrary triangulations and for a 
change of triangulation in time. Quantization of this framework is 
discussed further by H\"ohn\cite{PH1}.

Bonzom and Dittrich\cite{BD1} and then Asante, Dittrich and 
Haggard\cite{ADH1} used Regge calculus to calculate one-loop partition 
functions for gravity (for regions with boundary) and to define holographic 
duals for three-dimensional gravity and for the flat sector of 
four-dimensional gravity. It turns out that Regge calculus is well-suited 
to defining the partition function for regions with boundary and provides a 
regularization for computing the one-loop correction. A fully 
non-perturbative set-up and calculations using the Ponzano-Regge model are 
described by Dittrich, G\"oller, Livine and Riello\cite{DGLR}.

Although it is perhaps more relevant to the section on the Barrett-Crane model, it is worth mentioning the relation between Regge calculus and BF
theory obtained by Kisielowski\cite{MK}. A smooth manifold is obtained by
removing the hinges of a simplicial complex. The Regge geometry is then
encoded in a BF theory on the boundary of this manifold. The process
amounts to replacing the degrees of freedom of Regge calculus with discrete
degrees of freedom of topological BF theory. Bonzom\cite{VB} has also
explored the relation between BF theory and Regge calculus, showing that
together, the gluing relations and the simplicity constraints which turn BF
theory into simplicial gravity, contain the constraints of area-angle Regge
calculus\cite{BD}. The action includes the contribution from the Immirzi
parameter.

The possibility that the strength of gravitational interactions might
increase with distance has been explored in a series of papers by Hamber,
Toriumi and Williams. A set of effective field equations is formulated
incorporating the gravitational, vacuum-polarization induced running of
Newton's constant, G. This results in an accelerated power law expansion
for the Robertson-Walker universe, at times of the order of the inverse of
the Hubble constant\cite{HW1}. The implications for cosmological density
perturbations are also considered\cite{HT1}. The requirement of general
covariance for the effective field equations restricts the value of the
gravitational scaling exponent to be an integer greater than one\cite{HW2}.
The running of the cosmological constant, on the other hand, is shown in a
number of approaches to be inconsistent with general covariance\cite{HT2}.
To make the connection with experiment, Hamber\cite{HH} argues that the
lattice results suggest that the growth of Newton's constant, G, with
distance, should become observable only on very large distance scales,
comparable to the observed scaled cosmological constant. The hope is that
future high precision satellite experiments will be able to detect this
small quantum correction.

Quantum gravity in the limit of a large number of space-time dimensions has
been considered by Hamber and Williams\cite{HW3}. For a simplicial lattice
dual to a hypercube, a critical point was found, separating a weak coupling
from a strong coupling phase, where dominant contributions to the curvature
correlation functions were described by large closed random polygonal
surfaces.

Hamber and Williams\cite{HW4} have used the gravitational Wilson loop to
obtain information about the large-scale curvature properties of the
geometry. By comparing the resulting quantum averages to expected
semi-classical forms valid for macroscopic observers, it is possible to
identify the gravitational correlation length in the Wilson loop with the
observed large-scale curvature. The results imply a positive effective
cosmological constant at large distances.

Most of the quantum gravity work of Hamber and collaborators has used the
Euclidean lattice path integral approach, with numerical simulations using
Monte Carlo methods. To complement this, Hamber and Williams\cite{HW5} have
obtained a discrete form of the Wheeler-De Witt equation for the quantum
wave functional of the lattice. In the strong coupling limit, the wave
functional depends only on geometric quantities such as areas and volumes.
Explicit solutions in 2+1 dimensions are found\cite{HTW1}; a finite
correlation length emerges, that cuts off any infra-red divergences, and
there seems to be no weak-coupling perturbative phase. By contrast, in 3+1
dimensions\cite{HTW2}, the critical point in G has a non-zero value, but
the weak-coupling perturbative ground state appears to be
non-perturbatively unstable. The results obtained seem to suggest that the
Lorentzian and Euclidean formulations of lattice quantum gravity belong to
the same field-theoretic universality class.

In work related to that of Khavari\cite{PK}, using the causal structure in
Lorentzian space-time and the fact that certain simplices are not
constrained by the triangle inequalities, Tate and Visser\cite{TV2} set up
Lorentzian signature models of quantum Regge calculus, showing that these
are not related to Euclidean models by a simple Wick rotation. The lack of
the triangle inequality constraints means that it is easier to do
analytical calculations and that numerical simulations are more
computationally efficient. They set up the path integral for a model in 1+1
dimensions, obtaining scaling relations for the path integral and showing
that spikes are absent, then discuss the model in higher dimensions.

Causal dynamical triangulations are based on a Lorentzian simplicial
lattice and use the Regge action, but rather than integrating over the edge
lengths in the path integral, sums are performed over triangulations using
the Pachner moves. A sophisticated and detailed analysis of the phase
structure in 3+1 dimensions has been performed and we refer the reader to
reviews by Ambjorn, Loll and collaborators\cite{CDT}.

\section{The Barrett-Crane model}

\newcommand{\SU}{{\mathrm{SU}}}
\newcommand{\SO}{{\mathrm{SO}}}
\newcommand{\Spin}{{\mathrm{Spin}}}

The original development of the Barrett-Crane model is explained in 
detail in the earlier review\cite{RW}. The two versions of the model 
are based on representations of the group $\SO(4)$ in the Euclidean 
case or $\SO(3,1)$ in the Lorentzian case.  After the initial flurry
of excitement of the model as a potential four-dimensional quantum 
gravity theory, the hard work of establishing its properties continued 
at a slower pace. Different proposals for the normalisation factors on 
the lower-dimesional faces were examined, arriving at a reasonable 
scheme after numerical simulations of several proposals\cite{BCHT}.
The large-spin asymptotics of the amplitude of a single 4-simplex was 
established in a precise way by analytical means \cite{BS,FL} and 
confirmed by numerical means \cite{BCE}. This showed that although the 
desired four-dimensional `Regge calculus' geometries are present and 
contribute the Regge action to the amplitude, some degenerate 
configurations typically dominate. This presents a physical picture 
that is not so clear overall.

The issue of understanding the geometry of several 4-simplices glued 
together proves to be a hard problem. The main issue is that the data 
that is matched on a pair of 4-simplices is the areas of the four
common triangles, but this information is not enough to guarantee a 
continuous metric spanning the two. A geometrical interpretation of
this gluing is lacking, except perhaps across a null surface in the 
Lorentzian case \cite{JB}. 

The Pachner moves for the gluing have been analysed 
recently\cite{ADH} and it is argued there that the action of `area Regge 
calculus' (see section 2) provides a plausible 
semiclassical model for the phase of the Barrett-Crane amplitude. 
There is evidence that when more than two simplices are considered, 
the possible metric discontinuity is rather more restricted; for 
example along planes the discontinuity vanishes except in the null 
case \cite{WW}. There is much about the geometry of the Barrett-Crane 
model that is still to be understood.

Discontent with the gluing conditions led to some substantial 
extensions of the Barrett-Crane model. These new models were 
formulated by Freidel-Krasnov (FK) \cite{FK} and by 
Engle-Levine-Pereira-Rovelli (EPRL) \cite{EPRL}, and in the Lorentzian case 
by FK and Pereira\cite{FK,P}. The idea is to allow a vector space of 
intertwiners on each tetrahedron, so that the gluing between two 
4-simplices involves an inner product in this intertwiner space. Another closely related spin foam model for 4d quantum gravity based on the imposition of simplicity constraints via non-commutative flux variables was proposed, to deal with these and other issues, by Baratin-Oriti \cite{BO} (so far only in the Riemannian case). 

Representations of $\SO(4)$ are determined by a pair of half-integers 
$(j,j')$ which label the total spin of each factor in the spin 
covering $\Spin(4)\cong\SU(2)\times\SU(2)$. Choosing to sum over all 
representations $(j,j')$ on each triangle and the full intertwiner 
space on each tetrahedron leads back to the Ooguri model (or 
Crane-Yetter model in the $q$-deformed case), whereas limiting to the 
simple representations $(j,j)$ and the one-dimensional intertwiner 
space determined by the `canonical vertex' gives the Barrett-Crane 
model again. The idea of the new models is to take an intermediate 
case where $j'=cj$ for a global constant $c$, and on each tetrahedron 
a vector space of intertwiners for the group $\SU(2)$ (not $\SO(4)$). 
These intertwiners are supposed to give a more complete description 
of the quantum geometry of a tetrahedron. This means that these 
degrees of freedom can propagate into the neighbouring 4-simplex in 
a manifold, giving a more geometric gluing. 

The constant $c$ is determined by an Immirzi parameter, as introduced 
in loop quantum gravity, giving an asymmetry in the action between 
the self-dual and anti-self-dual parts of the curvature. For some 
values of $c$, the EPRL and FK models are identical, and for other 
values they are different but similar in construction.

As before, the first semiclassical analysis of the amplitudes for the 
new models was for the single 4-simplex, in the Euclidean 
case \cite{BDFGH} and in the Lorentzian case \cite{BDFHP}. These papers 
also provide a more precise definition of the models themselves. The 
boundary data is now more complicated because, beside the areas of the 
triangles, there is also data for the geometry of each tetrahedron. 
The amplitude for most boundary data is exponentially damped because 
the integral representation for the amplitude contains no stationary 
points. The interesting cases are where there are stationary points so 
that the amplitudes are not damped. The analysis shows that there is a 
class of such boundary data called `Regge-like' where all of the data 
is consistent with a flat metric in the 4-simplex. In terms of this 
metric, the phase of the 4-simplex amplitude is again the Regge action 
for a 4-simplex. There are however some other important
configurations, in the same way as in the original Barrett-Crane
model, but with different details. These other configurations are 
three-dimensional `vector geometries' or even lower-dimensional 
geometries, and do not have a Regge calculus interpretation.

Probably the simplest explanation of this asymptotics is given by 
considering the 4-simplex amplitudes as squares of 15j symbols and 
analysing the rather simpler asymptotics of the 15j 
symbols \cite{BFH}. Amazingly, the 15j symbols can also have 
Regge-like boundary data with asymptotics determined by the Regge 
calculus action, even though these symbols belong to the Ooguri model, 
which is not normally regarded as a gravity model at all.

Numerical methods to compute the 4-simplex amplitudes were developed for the 15j symbols \cite{Dona:2017dvf}, and  for the Lorentzian EPRL model \cite{Dona:2019dkf}, giving confirmations of the asymptotic formulae. These papers also derive a number of generalisations of the asymptotic formulae together with their geometric interpretations.

The EPRL and FK models still have difficulties with gluing. 
Considering a spatial slice of spacetime (i.e., a 3-manifold), each 
tetrahedron now has a quantum state space that can distinguish 
between the flat geometries with the same area for each triangle. But 
unfortunately, this data does not always glue together continuously 
from one tetrahedron to another across a common triangle. In essence, 
it is like the problem with the original Barrett-Crane model but in
one lower dimension. It is also similar to the gluing problem for the 
geometries associated to states in loop quantum gravity\cite{FS}.

Several works have analysed the EPRL/FK models when a number of 
4-simplices are glued together to form a manifold. Replacing each 
4-simplex amplitude with its asymptotic expression for large spins 
results in an amplitude for a triangulated manifold whose phase part 
is the Regge calculus action, but with additional independent 
variables\cite{HZE,HZL}. This procedure is only a heuristic one as 
the spin variables on interior triangles are summed over all values, 
so that it is not clear whether the asymptotic formula is a useful 
approximation. Nevertheless, continuing with this formula and looking 
for the stationary points when the spin is varied leads to the naive 
conclusion that the geometry is flat, at least in an asymptotic 
limit\cite{BZ}. More careful analyses are underway\cite{HelKam,O} but 
it is too soon for a definitive conclusion.


\section{Group field theories}

Group field theories (GFTs)\cite{GFT} bring Regge's intuition about 
describing space-time, geometry and gravity in terms of piecewise-flat, 
simplicial structures to a whole new level. They extend some of the 
simplicial quantum gravity approaches discussed above, merging them
into a single framework. They were first introduced\cite{boulatov} 
as an enrichment of tensor models, themselves a generalization of 
matrix models for two-dimensional gravity, to higher dimensions, and a 
generating functional for Euclidean dynamical triangulations. This enrichment 
defined instead a generating functional for sums over triangulations 
weighted by the state sum models for topological BF theory in three dimensions
(Ponzano-Regge model) and four dimensions, based on the representation 
theory of $SU(2)$. Then, group field theories were found\cite{DPFKR} 
to provide a generating functional for the Barrett-Crane model and a 
tentative definition of the dynamics of four-dimensional quantum
gravity, in the 
language of loop quantum gravity and spin foam models. Many
developments 
over the last twenty years form now a massive body of work making group 
field theories a promising formalism for a quantum gravity theory 
based on discrete structures. They have also led to a new perspective, 
in which spacetime and geometry are emergent notions, to be extracted 
from the collective behaviour of the discrete GFT structures, seen as 
fundamental entities\cite{emergence}. 

GFTs can be understood as quantum field theories for elementary
objects (the \lq quanta\rq  ~of the GFT field) corresponding to quantum 
tetrahedra. They rely on a description of simplicial geometry in terms 
of group- and representation-theoretic data\cite{baezbarrett}, also 
important in the context of spin foam models and loop quantum
gravity. The relevant phase spaces are cotangent bundles of group 
manifolds (usually $SU(2)$, $Spin(4)$ or $SL(2,\mathbb{C}$)) and the 
quantization leads to Hilbert spaces which admit a complete
orthonormal basis of spin network states. In fact, the fundamental 
tetrahedra (in four-dimensional models) can be seen as dual to spin network
vertices, with the outgoing links corresponding to the faces of the 
tetrahedra. The variables encoding the geometry of the simplicial 
structures are then group elements (corresponding to a discretized 
connection) and group representations, corresponding to eigenvalues 
of quantized face areas. A non-commutative description of the states 
can also be given \cite{flux}, in terms of Lie algebra elements 
corresponding to metric variables associated to the same faces. 
Generic quantum states are then many-body states formed by several 
quantum tetrahedra, including those in which the tetrahedra are glued 
to form extended three-dimensional triangulations. A Fock space
description of the 
GFT Hilbert space has been given by Oriti\cite{GFTfock}, clarifying also 
the nature of these models as second quantized descriptions of loop 
quantum gravity-type theories.

On these kinematical states, different dynamical prescriptions can be 
imposed. They are encoded in a choice of GFT action for a field whose 
arguments are the same data characterizing the geometry of a
tetrahedron.  The defining feature is a non-local pairing of field 
arguments in the interaction terms, reflecting the gluing of the 
tetrahedra to form the boundary of four-dimensional cells, taken as
the fundamental 
building block of their interaction processes. The perturbative 
expansion of the GFT partition function defines a sum over Feynman 
diagrams dual to four-dimensional cellular complexes, with the 
interaction vertices 
defining the constituting cells and the propagator enforcing their 
gluing. One route towards model building\cite{tensor} is based on the 
fact that GFT fields can be seen as generalised tensors with indices 
on the group manifold, transforming under unitary groups acting on the 
same indices. Thus, a theory space can be defined including all 
possible unitary invariants as interaction terms. Such tensorial GFTs 
can then take advantage directly of the many results obtained  in the 
context of tensor models\cite{tensor}, concerning the control over 
the topology of the GFT interaction diagrams, and their statistical 
analysis and critical behaviour. In fact, when the Lie group defining 
the domain of GFT fields is replaced by a finite group or simply by 
some finite set, then GFT models become tensor models, and their 
Feynman amplitudes depend only on the combinatorics of the underlying 
diagram; they actually become akin to those of the dynamical 
triangulations approach. Beside this tensorial setting, there have
been few analyses of symmetries in GFT models\cite{GFTsymmetries}, 
and thus there is little control over possible theory spaces. Model 
building has been based, then, on the choice of having Feynman
diagrams corresponding to triangulations (which suggests interactions 
with the combinatorics of 4-simplices) and on the desire to have the 
corresponding Feynman amplitudes coinciding with the most promising 
spin foam models, proposed in the context of simplicial quantum 
gravity and loop quantum gravity\cite{DPFKR, GFT, BO},and inspired 
by the formulation of (continuum and discrete) GR as a constrained 
BF theory. The latter requirement translates into specific choices of 
kinetic and interaction terms in the GFT action, for each desired 
model. In fact, one can show that the correspondence between GFTs 
and spin foam models is generic\cite{mikecarlo}: for any given spin 
foam model defined on a given cellular complex, there is a GFT model 
whose Feynman amplitudes coincide with it on the Feynman diagram dual 
to the same complex (and vice versa, any given GFT model defines a 
corresponding spin foam model, in its perturbative expansion). 
Similarly generic is the fact that, when one formulates the same 
GFT models in terms of Lie algebra variables, the corresponding 
Feynman amplitudes take the form of (non-commutative) simplicial 
gravity path integrals in first order variables, on the same complex\cite{BO}. 

From the point of view of simplicial quantum gravity, therefore, group 
field theories define in their perturbative expansion a quantum
dynamics combining a sum over discrete geometric data corresponding 
to a first order version of quantum Regge calculus, associated to a
given lattice, with a  sum over lattices, including very refined ones,  
in the spirit of dynamical triangulations. The problem becomes, of 
course, to show the consistency of such quantum theory, i.e. to 
control analytically the combined sum, and to \lq resum it\rq, i.e. 
to define the full partition function, thus the continuum limit. 
This is where the quantum field theory nature of GFTs becomes 
crucial. Both problems, in fact, become problems in the 
renormalization of GFT models: to prove their perturbative 
renormalizability, and to define their full RG flow and continuum 
phase diagram. This has become a very active research direction 
over the last ten years\cite{sylvain}, in parallel with the 
developments in tensor models. Among the many results, it has led 
to: a better understanding of the divergences of four-dimensional quantum 
gravity models for constrained BF theories \cite{SFdivergences}, 
and a complete characterization of them for topological BF 
models \cite{matteovalentin}; the rigorous proof of 
renormalizability of several tensorial GFT models \cite{CRO}, 
both abelian and non-abelian, in several dimensions; constructive 
analyses of the same tensorial GFT models\cite{GFTconstructive}; 
the application of functional renormalization group techniques 
with a first characterization of their flows and phase diagrams 
(in simple truncations), confirming their asymptotic freedom or 
safety, and providing hints of new phases\cite{FRG-GFT}. A first 
analysis of inequivalent (coherent) representations of GFT 
observables algebras, possibly corresponding to different phases 
of GFT systems, has been done by Kegeles, Oriti and Tomlin\cite{alex}.

The extraction of effective physics from GFT models can be pursued 
using methods from simplicial (quantum) gravity and spin foam models, 
at the level of GFT perturbation theory. The generating functional 
for superpositions of discrete structures defined by GFT models, as 
well as their second quantized formalism, are useful tools to do so. 
They also offer, however, new avenues to study emergent continuum 
physics bypassing to some extent the discrete gravity picture, and 
aiming at coarse-grained information encoding many discrete gravity 
degrees of freedom. In one such line of research\cite{GFTcosmoReview},
effective cosmological physics is extracted 
from the hydrodynamics of GFT models and more precisely from the 
mean field description corresponding to simple GFT condensate states. 
Among the many results, we cite the evidence that (within this 
hydrodynamic approximation): a homogeneous and isotropic  universe 
satisfies a modified Friedmann equation with the correct classical 
limit but with a bouncing regime replacing the classical big bang 
singularity\cite{GFTbouncing}, similar to what is found in loop 
quantum cosmology; GFT interactions can modify this dynamics to 
give an extended period of acceleration following the bounce (a 
sort of quantum gravity-driven inflation), without the need to 
introduce additional (inflaton-like) degrees of freedom, and a 
later recollapse and cyclic evolution\cite{GFTaccelerated}; the 
extension of the formalism to include 
anisotropies\cite{GFTanisotropies} and 
inhomogeneities\cite{GFTperturbations}, with approximate scale 
invariance found in the spectrum of cosmological perturbations.

\section{Conclusions}
Applying the geometric concept of piecewise linear spaces to general
relativity, Regge provided the basis for very fruitful reseach in a
number of areas, including calculations of the classical evolution of
model universes and formulations of quantum gravity. Since the review
paper of 2000, in particular there has been outstanding progress in
the development
of quantum gravity models, which arise directly from Regge's work. These
have the potential to provide a consistent theory of 
quantum gravity with wide-ranging implications in physics. 

\subsection*{Acknowledgements}
The work of RMW has been partially supported by STFC consolidated
grant ST/P000681/1.

\addtocontents{toc}   
{\contentsline{chapter}{\numberline{}{\protect\\[20pt]\bf\hfill Contents final formatting will be taken care by the publisher\hfill}}{}}

\end{document}